\documentclass[aps,pre,reprint,showpacs,amsmath,amssymb,floatfix,10pt]{revtex4-1} 

\usepackage[pdftex]{graphicx}       
\usepackage{txfonts}        
\usepackage{amsmath}
\usepackage{amssymb}
\usepackage{mathtools}
\usepackage{mathrsfs}	

\newcommand{\scT}{\mathcal{T}}

\newcommand{\kcinf}{\kappa_\text{a}}
\newcommand{\Tcn}{T_\text{a}^{(n)}}
\newcommand{\Tcinf}{T_\text{a}}

\newcommand{\RRp}{R^2_{\perp, n}}
\newcommand{\RRpll}{R^2_{\parallel,n}}

\newcommand{\eref}[1]{Eq.~\eqref{#1}}
\newcommand{\fref}[1]{Fig.~\ref{#1}}
\newcommand{\sref}[1]{Section \ref{#1}}

\usepackage{hyperref} 
\hypersetup{colorlinks,citecolor=blue,filecolor=blue,linkcolor=blue,urlcolor=blue}

\begin{document}

\title{Universality of crossover scaling for the adsorption transition of lattice polymers}

\author{C. J. Bradly} \email{chris.bradly@unimelb.edu.au}
\author{A. L. Owczarek}\email{owczarek@unimelb.edu.au}
\affiliation{School of Mathematics and Statistics, University of Melbourne, Victoria 3010, Australia}
\author{T. Prellberg} \email{t.prellberg@qmul.ac.uk}
\affiliation{School of Mathematical Sciences, Queen Mary University of
  London, Mile End Road, London, E1 4NS, United Kingdom}
\date{\today}

\begin{abstract}
Recently, it has been proposed that the adsorption transition for a single polymer in dilute solution, modeled by lattice walks in three dimensions, is not universal with respect to inter-monomer interactions. Moreover, it has been conjectured that key critical exponents $\phi$, measuring the growth of the contacts with the surface at the adsorption point, and $1/\delta$, which measures the finite-size shift of the critical temperature, are not the same.  However, applying standard scaling arguments the two key critical exponents should rather be identical, hence pointing to a potential breakdown of these standard scaling arguments.
Both of these conjectures are in contrast to the well studied situation in two dimensions, where there are exact results from conformal field theory: these exponents are both accepted to be $1/2$ and universal.

We use the flatPERM algorithm to simulate self-avoiding walks and trails on the hexagonal, square and simple cubic lattices up to length $1024$ to investigate these claims. Walks can be seen as a repulsive limit of inter-monomer interaction for trails, allowing us to probe the universality of adsorption. For each lattice model we analyze several thermodynamic properties to produce different methods of estimating the critical temperature and the key exponents. We test our methodology on the two-dimensional cases and the resulting spread in values for $\phi$ and $1/\delta$ indicates that there is a systematic error which can far exceed the statistical error usually reported. We further suggest a methodology for consistent estimation of the key adsorption exponents which gives $\phi=1/\delta=0.484(4)$ in three dimensions. Hence, we conclude that in three dimensions these critical exponents indeed differ from the mean-field value of $1/2$, as had previously been calculated, but cannot find evidence that they differ from each other. Importantly, we also find \emph{no substantive evidence} of any non-universality in the polymer adsorption transition.

\end{abstract}

\maketitle

\section{Introduction}
\label{sec:Intro}

The adsorption of single polymers in dilute solution onto a substrate has been extensively studied for many years via a variety of theoretical models and techniques \cite{Eisenriegler1982,DeBell1993,Vrbova1996,Vrbova1998,Vrbova1999,Grassberger2005,Owczarek2007,Luo2008,Klushin2013,Plascak2017}. The critical phenomenon associated with this transition is a fundamental one in the landscape of statistical physics. 
In dilute solutions at high temperatures the configuration of the polymer is dominated by entropic repulsion, forming an expanded phase where the polymer is desorbed from the surface. 
Of particular interest is when there is also an attractive interaction between the monomers and the surface. In this situation, the configuration of the polymer is further influenced by energetic considerations and at low temperatures the polymer seeks to lower its energy by staying close to the surface and is adsorbed. The transition between these regions occurs at the adsorption temperature $\Tcinf$ where the polymers display critical phenomena \cite{DeBell1993}. 
Many generalizations have been studied and aspects of this behavior still attract much interest \cite{Grassberger2005,Luo2008,Klushin2013,Plascak2017}. One fruitful set of models use self-avoiding paths on a lattice to represent the polymer.

If we consider the thermodynamic limit of infinitely long polymers, the internal energy per monomer $u_\infty$ associated with contacts with a surface is expected to be zero for temperatures above $\Tcinf$ and strictly positive below $\Tcinf$. The singular behavior for $T \rightarrow \Tcinf^-$ is given by the thermal exponent $\alpha$
\begin{equation}
	u_\infty \sim \left(\Tcinf-T\right)^{1-\alpha}\;,
	\label{eq:alpha}
\end{equation}
while the length scaling behavior of the finite length internal energy $u_n$ per monomer
defines an exponent usually labeled $\phi$
\begin{equation}
	u_n  = \frac{\langle m \rangle}{n} \sim n^{\phi-1}\;,
	\label{eq:phi}
\end{equation}
where $\langle m \rangle$ is the mean number of interactions (contacts with the surface).
This scaling implies that at $T_a$ there is
$
\langle m \rangle \sim n^\phi .
$
For high temperatures $\langle m \rangle$ is expected to be bounded while at low temperatures $\langle m \rangle$ is asymptotically linear in length $n$ so that a positive thermodynamic internal energy exists. This broad behavior characterizes the adsorption transition. Now the upper critical dimension for the adsorption transition is expected to be $d_\text{u}=4$ and the mean field value of $\phi$ is $1/2$. Interestingly, in two dimensions exact results from both directed models and the hexagonal lattice predict that $\phi=1/2$. Careful simulations in three dimensions \cite{Grassberger2005} have verified the prediction of field theoretic expansions around $d=d_\text{u}=4$ that $\phi\neq 1/2$ in three dimensions. A value just below $1/2$ was estimated by Grassberger as $0.484(3)$ \cite{Grassberger2005}. 

One can also consider the scaling around the adsorption point in temperature and length together. We denote the exponent controlling the crossover to be $1/\delta$ in line with previous works. Until recently it was accepted that \smash{$1/\delta=\phi$} (we detail below one scaling argument for this correspondence). In fact, in both mean field theory and in two dimensions \smash{$1/\delta=1/2$}. Luo \cite{Luo2008} suggested that in three dimensions they may be different. Recently, it was further suggested by Plascak {\em et al.}~\cite{Plascak2017}, that both exponents may not be universal: to be specific, by adding monomer-monomer interactions to the model both these exponents depend continuously on the strength of the interaction even well away from any critical point induced by the  monomer-monomer interactions. It is well known that when monomer-monomer interactions are sufficiently positive (low temperatures) a collapse transition can occur. They suggested that even repulsive interactions can induce a a non-universality. 

To investigate the numerical validity of these claims we have simulated a range of models in both two and three dimensions. We consider self-avoiding walks (SAWs) on the hexagonal, square and simple cubic lattices, and self-avoiding trails (SATs) on the square and simple cubic lattices. We do not consider monomer-monomer interactions in the model, in which case SAWs and SATs are believed to be in the same universality class in all dimensions with the same finite-size scaling exponents. 
Although well studied, we include the square and hexagonal lattice models as a useful benchmark for our methods since there is little dispute about the adsorption transition scaling in two dimensions. In particular, the case of self-avoiding walks on the hexgonal lattice has been solved  and the critical exponents, transition temperature and connective constant are known exactly \cite{Batchelor1995}.

For all of these lattice models, we use a variety of methods of analysis designed to estimate the key critical exponents including those used by Plascak {\em et al.}~\cite{Plascak2017}. Even in the two-dimensional lattice models it is apparent that the systematic error inherent in all these methods often swamps the statistical error. Moreover, the spread of the results give a much better correlation with the correct values for the critical temperature and exponents \smash{$\phi=1/\delta=1/2$} than any individual estimate. With this in mind, we find that in the three-dimensional case the central estimates agree with Grassberger's estimate \cite{Grassberger2005} that \smash{$\phi<1/2$}. However, we find no evidence that $1/\delta$ and $\phi$ are different as suggested by Luo \cite{Luo2008} and Plascak {\em et al.}~\cite{Plascak2017}. 
Moreover,  we find the values for SAWs and SATs to be numerically equivalent and so find no evidence of any non-universality as suggested by Plascak {\em et al.}~\cite{Plascak2017}. We rather suggest that the previous results were simply a case of systematic errors from higher order corrections to scaling leading to apparent differences. We finally provide our own estimate of \smash{$\phi=1/\delta=0.484(4)$} in three dimensions.

\section{The models}
\label{sec:Model}

\begin{figure}[t!]
	\centering
	\includegraphics[width=\columnwidth]{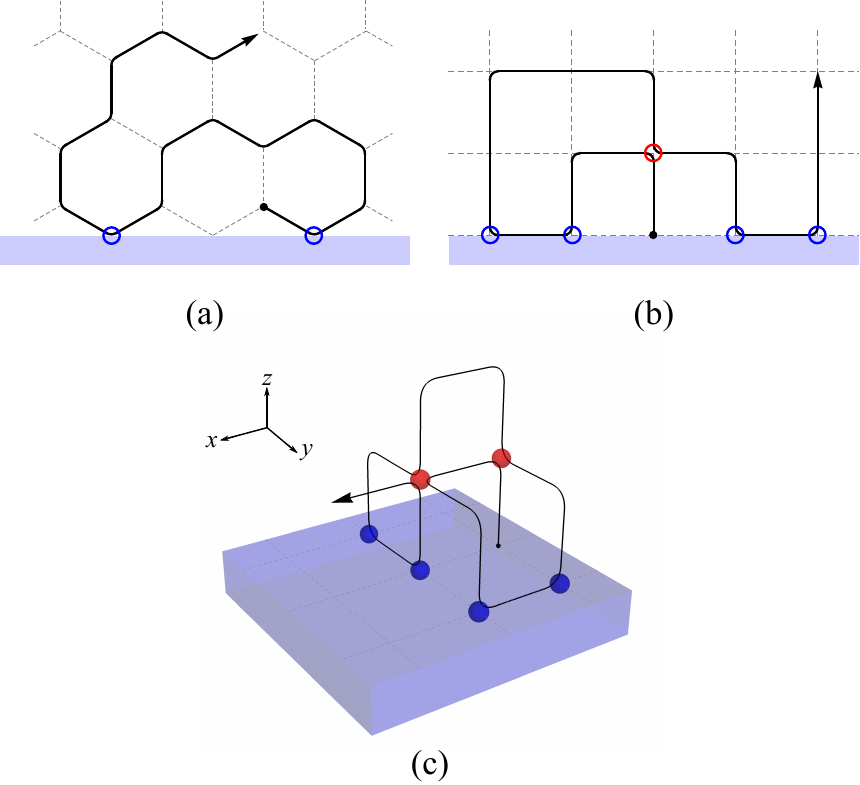}
	\caption{Self-avoiding walk on (a) the hexagonal lattice and self-avoiding trails on (b) the square lattice and (c) the simple cubic lattice, in the presence of an impermeable adsorbing surface. Sites in contact with the surface, other than the origin, are marked blue. Walks on the square and simple cubic lattices are the same with respect to the surface, but multiply-visited sites, marked red,  are forbidden. }
	\vspace{-0.5cm}
	\label{fig:LatticePaths}
\end{figure}

A self-avoiding trail (SAT) is a lattice path with the restriction that no two bonds between consecutive steps may overlap. A self-avoiding walk (SAW) has the additional restriction that lattice sites cannot be occupied more than once. The set of SAWs is a subset of the set of SATs.

The impermeable adsorbing surface is represented by restricting trails/walks to \smash{$x_d\ge 0$} for a $d$-dimensional lattice with coordinate system $x_i$, for \smash{$i=1,\ldots d$}. \fref{fig:LatticePaths} shows fragments of a SAW on (a) the hexagonal lattice and a SAT on (b) the square lattice and (c) the simple cubic lattice, near an impermeable boundary layer. In particular, note that on the hexagonal lattice, only every second site is considered on the surface.
The surface-monomer interaction is modeled by assigning an energy $-\epsilon$ to any monomer on the surface \smash{$x_d = 0$}. This does not include the initial point at the origin fixing the path to the surface.

\subsection{Thermodynamic quantities}
\label{sec:Thermo}

A trail (or walk) $\psi_n$ of length $n$ with one end fixed to the surface and with $m$ contacts with that surface has total interaction energy $-m\epsilon$ and corresponding Boltzmann weight $\kappa^m$, where \smash{$\kappa = \exp(\epsilon/k_\text{B}T)$}. Thus, the partition function of the set $\scT_n$ of walks/trails of length $n$ is
\begin{equation}
    Z_n(\kappa) = \sum_{\psi_n \in \scT_n} \kappa^m.
    \label{eq:Partition}
\end{equation}
The (reduced) finite-size free energy is
\begin{equation}
	f_n(\kappa)  = - \frac{1}{n} \log Z_n(\kappa) ,
\end{equation}
while the thermodynamic limit is given by
\begin{equation}
	f_\infty(\kappa) = \lim_{n\rightarrow\infty} f_n(\kappa).
\end{equation}
A general thermodynamic quantity is
\begin{equation}
    \langle Q \rangle(\kappa) = \frac{1}{Z_n(\kappa)}\sum_{\psi_n \in \scT_n} \kappa^m Q(\psi_n).
    \label{eq:ThermoQuantity}
\end{equation}
In particular, we are interested in the internal energy
\begin{equation}
    u_n (\kappa) = \frac{\langle m \rangle}{n},
    \label{eq:InternalEnergy}
\end{equation}
which, considered as the fraction of the walk/trail that is adsorbed to the surface, serves as our order parameter. 

The other quantity of interest is the mean-squared end-to-end radius $ R^2_n$. In the presence of an interacting surface we distinguish between the parallel and perpendicular components, with respect to the surface. For a $d$-dimensional system these components are defined as
\begin{alignat}{1}
    \RRpll (\kappa) &= \sum_{i=1}^{d-1}\langle {x_{i,n}}^2 \rangle, \\
    \RRp (\kappa) &= \langle x_{d,n}^2 \rangle, \\
    \label{eq:EndToEndRadius}
\end{alignat}
where $x_{i,n}$ is the $i$-th coordinate of the $n$-step of the path.
Recall that for the simple cubic lattice the adsorbing surface is the $(x_1,x_2)$-plane at \smash{$x_3=0$} and in two dimensions the surface is the \smash{$x_1$-axis} at \smash{$x_2=0$}.

\subsection{Scaling laws and critical temperatures}
\label{sec:Scaling}

The exponent $\phi$, usually expected to be universal, determines the scaling of the order parameter at the critical point for long chains: $u_n\sim n^{\phi-1}$. For the finite values of $n$ considered in numerical simulations, it is necessary to also include finite-size correction terms. From finite-size scaling theory we have
\begin{equation}
    u_n \sim n^{\phi-1} f_u^\text{(0)}(x) [1 + n^{-\Delta}f_u^\text{(1)}(x) +\ldots ],
    \label{eq:UnScaling}
\end{equation}
where the $f^{(i)}$ are  finite-size scaling functions of the scaling variable \smash{$x=(\Tcinf-T)\,n^{1/\delta}$} and \smash{$\Delta\lesssim 1$} is the first correction-to-scaling term. 
The exponent $1/\delta$ therefore describes the \emph{crossover} around the adsorption critical point. It can also be described as  the shift exponent associated with the deviation of temperature from the critical point. That is, the finite-length critical temperature differs from the infinite-length critical temperature according to
\begin{equation}
    \Tcn \sim \Tcinf + n^{-1/\delta} f_T^\text{(0)}(x) [1 + n^{-\Delta}  f_T^\text{(1)}(x)+\ldots].
    \label{eq:TempScaling}
\end{equation}

Somewhat confusingly in the literature, the exponent $\phi$ is often referred to as the \emph{crossover} exponent since it has, until recently, been accepted that there is a crossover scaling variable \smash{$x=(\Tcinf-T)\,n^{\phi}$} describing the scaling around the adsorption point. Below we provide a scaling argument that connects $\phi$ and $1/\delta$  \cite{Eisenriegler1982,Rensburg2004}. The argument starts with the scaling of the partition function. At any fixed temperature the partition function scales as
\begin{equation}
	Z_n(\kappa) \sim A \mu^n n^{\gamma^{(1)} -1},
\end{equation}
where $\gamma^{(1)}$ is the entropic exponent that takes on one value at high temperatures and different values at the adsorption point and at low temperatures. Let us denote the value at the adsorption point as $\gamma^{(1)}_\text{a}$. The connective constant $\mu(\kappa) = \log f_\infty^{-1} $ is temperature dependent and directly related to the thermodynamic limit of the free energy. Following the same standard scaling hypothesis as above, one expects
\begin{equation}
	Z_n(\kappa) \sim A \; \mu_\text{a}^n \; n^{\gamma^{(1)} _\text{a} -1}\, {\cal Z}\left( t n^{1/\delta} \right),
\end{equation}
for $\kappa$ near $\kcinf$, where $\mu_\text{a} =\mu(\kcinf)$ and $t=T_a-T$.
This form can be deduced from a similar ansatz for the scaling of the corresponding generating function. The (reduced) finite-size free energy therefore scales as
\begin{equation}
	f_n(\kappa)  \sim -\frac{1}{n} \log\left( A n^{\gamma^{(1)} _a -1} \right) +   f_\infty(\kcinf) + \frac{1}{n} {\cal F}\left( t n^{1/\delta} \right),
\end{equation}
where the first terms are temperature independent.
The key point is that the internal energy is given, up to a multiplicative constant, by the temperature derivative of the free energy, so  this form immediately implies that 
\begin{equation}
    u_n \sim   n^{1/\delta-1} {\cal F^\prime}\left( t n^{1/\delta} \right).
	\label{eq:ScalingArg}
\end{equation}
Comparing \eref{eq:ScalingArg} to \eref{eq:UnScaling} yields $\phi=1/\delta \,$.

A related argument concerns the crossover from the temperature scaling of the internal energy in \eref{eq:alpha} to the length scaling in \eref{eq:phi} via the crossover form in \eref{eq:UnScaling}. The scaling function should behave as
\begin{equation}
	f_u^\text{(0)}(x) \sim x^{(1-\phi)\delta},
\end{equation}
which eliminates the length dependence and leads to
\begin{equation}
	1- \alpha  =  (1-\phi)\delta.
\end{equation}
If we also accept the previous argument that $\phi=1/\delta$ then this implies that
\begin{equation}
	\alpha  =  2 -\delta = 2 - \frac{1}{\phi}\;.
\end{equation}

\begin{figure*}[t!]
	\centering
	\includegraphics[width=\textwidth]{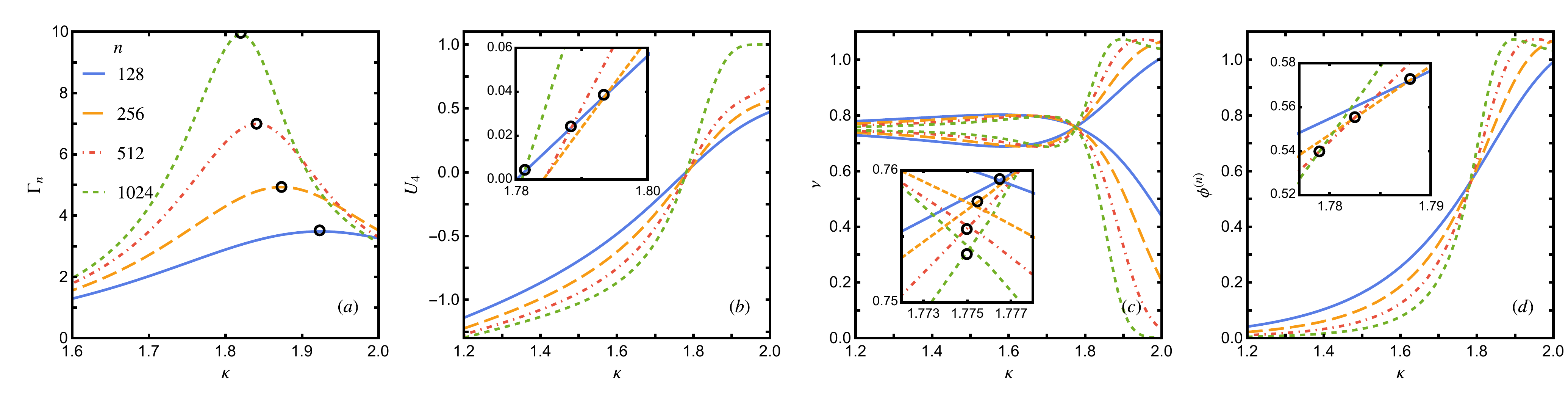}
	\vspace{-1.cm}
	\caption{The four methods for obtaining $\Tcn$, illustrated with data for SAWs on the square lattice. For clarity, error bars of thermodynamic quantities have been omitted and only \smash{$n=128,256,512,1024$} are shown. Black circles mark (a) $\Gamma$: positions of $\max\Gamma_n$, (b) BC: intersections of $U_4$ at various $n$ with $U_4$ at \smash{$n=128$}, (c) R2: intersections of $R^2$ exponents $\nu_\perp$ with $\nu_\parallel$, and (d) ratio: intersections of $\phi^{(n_i)}$ with $\phi^{(n_{i+1})}$. }
	\label{fig:TcnMethods}%
	\vspace{-0.5cm}
\end{figure*}

Despite these arguments, Luo \cite{Luo2008} conjectured that $\phi$ and $1/\delta$ may be different in three dimensions.
One way to extract $1/\delta$ separately rather than by calculating the temperature shift directly is to consider the log-derivative of $u_n$,
\begin{equation}
    \Gamma_n(\kappa) = \frac{d\log u_n}{dT} =
	(\log\kappa)^2
	\frac{\langle m^2 \rangle - \langle m \rangle^2}{\langle m \rangle}.
    \label{eq:LogDerivative}
\end{equation}
As a second derivative of the free energy, we expect a critical scaling form
\begin{equation}
    \max \Gamma_n \sim n^{1/\delta} f_\Gamma^\text{(0)}(x) [1 + n^{-\Delta} f_\Gamma^\text{(1)}(x) +\ldots].
    \label{eq:GammaScaling}
\end{equation}
By \eref{eq:LogDerivative}, $\Gamma_n$ is related to the specific heat. The peaks of the specific heat are often used to locate the collapse transition of trails in the bulk but this approach is inaccurate for locating the adsorption transition \cite{Rensburg2004}. Nevertheless, it is usually assumed that $x$ is small enough to use \eref{eq:GammaScaling} to determine $1/\delta$.

\section{Methods}
\label{sec:Methods}

The key to estimating $\phi$ and $1/\delta$ is to accurately locate the finite-size critical temperatures $\Tcn$. We explore four methods of calculating $\Tcn$, illustrated in \fref{fig:TcnMethods} using data for SAWs on a square lattice as an example. First, the simplest but least accurate is to consider the locations of $\max\Gamma_n$ as estimates of $\Tcn$; this method is labeled `$\Gamma$'. Despite the issues relating to the specific heat, it is a useful comparison to the other methods.

Second, we calculate the Binder cumulant
\begin{equation}
    U_4(\kappa) = 1 - \frac{1}{3}\frac{\langle m^4 \rangle}{\langle m^2 \rangle^2},
    \label{eq:Binder}
\end{equation}
a quantity that, for large $n$, tends toward a universal constant value at the critical point \cite{Binder1981}. Thus, intersections of curves of $U_4$ at different $n$ with the curve at fixed \smash{$n_\text{min}=128$} are used to locate the finite-size critical temperatures. This method is labeled `BC'.

The third method, labeled `R2', looks at the scaling of each component of the mean-squared end-to-end radius. For either component $i$,
\begin{equation}
    R_{i,n}^2 \sim n^{2\nu_i},
    \label{eq:R2Scaling}
\end{equation}
where \smash{$i=\perp,\parallel$} and the Flory exponent $\nu_i$ depends on the phase and dimension of the system and is calculated by simply inverting \eref{eq:R2Scaling}:
\begin{equation}
   \nu_i = \frac{1}{2} \log_2\frac{R_{i,n}^2}{R_{i,n/2}^2}.
   \label{eq:NuRatio}
\end{equation}
At high temperatures, the polymers are desorbed and both perpendicular and transverse components of $R^2$ scale as per the $d$-dimensional bulk. Below the adsorption temperature, the polymers' extent away from the surface vanishes and thus \smash{$R^2_\perp\to 0$} (or \smash{$\nu_\perp\to 0$}). The polymers are adsorbed to the surface to become a quasi-\smash{($d-1$)}-dimensional system and \smash{$\nu_\parallel^{(d)}\to \nu_\text{bulk}^{(d-1)}$}. At some intermediate temperature the components of $\nu$ cross and in fact the intersections locate the finite-size critical temperatures $\Tcn$. 

In view of \eref{eq:NuRatio}, the fourth method, labeled 'ratio', is to calculate the exponent $\phi$ directly as the leading order of the order parameter. That is, 
\begin{equation}
   \phi = 1 + \log_2\frac{u_{n}}{u_{n/2}}
   \label{eq:PhiRatio}
\end{equation}
is calculated over a range of $n$. As a function of temperature, it is known that, in addition to value of $1/2$ at the critical point, the scaling exponent of the internal energy vanishes at high temperatures and tends to unity at low temperature. For SAWs on a square lattice this is borne out in \fref{fig:TcnMethods}(d). Then, as with the R2 and BC methods, we can locate the critical temperatures $\Tcn$ from the intersections of curves of \eref{eq:PhiRatio} for successive values of $\{n_i,n_i+1\}$.

While finite-size scaling methods are the main focus, we can also consider other ways of estimating exponents. To that end, we consider that as well as the intersections for the ratio method locating the critical temperatures, \eref{eq:PhiRatio} is a direct estimate of $\phi$. This `direct' method provides a set of finite-size estimates, $\phi^{(n)}$, which, in the limit \smash{$n\to\infty$}, extrapolate to an alternative estimate of $\phi$ without reference to the scaling form \eref{eq:UnScaling} and its dependence on locating the critical temperatures.

\subsection{Numerical simulation}
\label{sec:Numerical}

Trails and walks are simulated using the flatPERM algorithm \cite{Prellberg2004}, an extension of the pruned and enriched Rosenbluth method (PERM) \cite{Grassberger1997}. The simulation works by growing a walk/trail on a given lattice up to some maximum length $N_\text{max}$. At each step the cumulative Rosenbluth \& Rosenbluth weight \cite{Rosenbluth1955} of the walk/trail is compared with the current estimate of the density of states $W_{n,m}$. If the current state has relatively low weight (i.e.~by being trapped or reaching the maximum length) the walk/trail is `pruned' back to an earlier state. On the other hand, if the current state has relatively high weight, then microcanonical quantities are measured and $W_{n,m}$ is updated. The state is then `enriched' by branching the simulation into several possible further paths (which are explored when the current path is eventually pruned back). When all branches are pruned a new iteration is started from the origin.

FlatPERM enhances this method by altering the prune/enrich choice such that the sample histogram is flat in the microcanonical parameters $n$ and $m$. Further improvements are made to account for the correlation between branches that are grown from the same enrichment point, which provides an estimate of the number of effectively independent samples. We also run 10 completely independent simulations for each case to estimate the statistical error.

The main output of the simulation is the density of states $W_{n,m}$ of walks/trails of length $n$ with $m$ contacts with the surface, for all \smash{$n\le N_\text{max}$}. Thermodynamic quantities are then given by the weighted sum
\begin{equation}
    \langle Q \rangle(\kappa) = \frac{\sum_{m} Q_m\kappa^m W_{n,m}}{\sum_{m} \kappa^m W_{n,m}}.
    \label{eq:FPQuantity}
\end{equation}
For example, the $q^\text{th}$ order moments needed for the thermodynamic quantities in \sref{sec:Thermo} are calculated directly as
\begin{equation}
    \langle m^q \rangle = \frac{\sum_{m=0}^n m^q\kappa^m W_{n,m}}{\sum_{m=0}^n \kappa^m W_{n,m}}.
    \label{eq:FPEnergy}
\end{equation}
Other microcanonical quantities $r_{\perp,n}^2$ and $r_{\parallel,n}^2$ are also calculated during the simulation.

\setlength{\tabcolsep}{4pt}
\begin{table}[t!]
	\caption{Details of flatPERM simulations. In all cases the number of samples and effectively independent samples is the average of 10 independent runs. 
	}
	\begin{tabular}{llrrrr}
	\hline \hline
	& Walks/	& \multicolumn{1}{l}{Max} &  &	\multicolumn{1}{l}{Samples at} & \multicolumn{1}{l}{Ind.~samples} \\ [-1pt]
	Lattice & trails & \multicolumn{1}{l}{length} & \multicolumn{1}{l}{Iterations }&  \multicolumn{1}{l}{max length}	& \multicolumn{1}{l}{max length}	\\ \hline
	hex & SAW & 4096 & $1.8\times 10^7$ & $2.3\times 10^9$ 		& $1.0\times 10^7$ \\ 
	hex & SAW & 1024 & $5.5\times 10^5$ & $2.0\times 10^{10}$ 	& $2.6\times 10^8$ \\[1ex]
	squ	& SAW & 1024 & $3.7\times 10^5$ & $3.9\times 10^{10}$ 	& $3.2\times 10^{8}$ \\
	squ	& SAT & 1024 & $3.7\times 10^5$ & $3.9\times 10^{10}$ 	& $3.1\times 10^{8}$ \\[1ex]
	sc	& SAW & 1024 & $4.4\times 10^5$ & $3.5\times 10^{10}$ & $5.4\times 10^8$ \\
	sc	& SAT & 1024 & $4.4\times 10^5$ & $3.4 \times 10^{10}$ & $5.9\times 10^8$ \\ 
	\hline \hline
	\end{tabular}
	\label{tab:SimDetails}
\end{table}

In this work we used the flatPERM algorithm to simulate walks and trails on the square and simple cubic lattices up to length $1024$, and walks on the hexagonal lattice at the exact adsorption transition, \smash{$\kcinf=1+\sqrt{2}$}, up to length $4096$ and without fixed weight up to length $1024$. Details of the simulations run in this work are summarized in Table \ref{tab:SimDetails}. Note that flatPERM is generally an athermal simulation but in the case of walks on the hexagonal lattice at the exact critical temperature, a fixed weight $\kcinf$ is applied at each step by altering the usual Rosenbluth \& Rosenbluth weight
. That is, the term $\kappa^m W_{n,m}$ in \eref{eq:FPQuantity} is calculated during the simulation (at fixed \smash{$\kappa=\kcinf$}) and the density of states is output as $W_n$; the sample histograms are {\em not} flattened with respect to $m$. This both saves memory and reduces equilibration time so that longer lengths can be simulated.

\section{Results and Discussion}
\label{sec:Results}
To understand the analysis, we will look at the case of SAWs on a square lattice in some detail before presenting the combined results for all lattices. First, some general remarks that apply to all cases. In all finite-size scaling fits, we assume that the scaling variable $x$ is constant with respect to $n$ so that the $f^{(i)}(x)$ may be treated as constants. This is readily verified to be true, although $x$ is not necessarily small in all cases. We find that the correction-to-scaling term is always necessary for a good fit and after considering the case of square SAWs we do not report the power-law only results. Finally, even with a correction-to-scaling term, we always consider \smash{$n=128,\ldots,1024$} since \smash{$n<100$} is too far from the scaling regime. 

\subsection{SAWs on square lattice}
\label{sec:ResultsSquSAWs}

\begin{figure}[t!]
	\centering
	\includegraphics[width=\columnwidth]{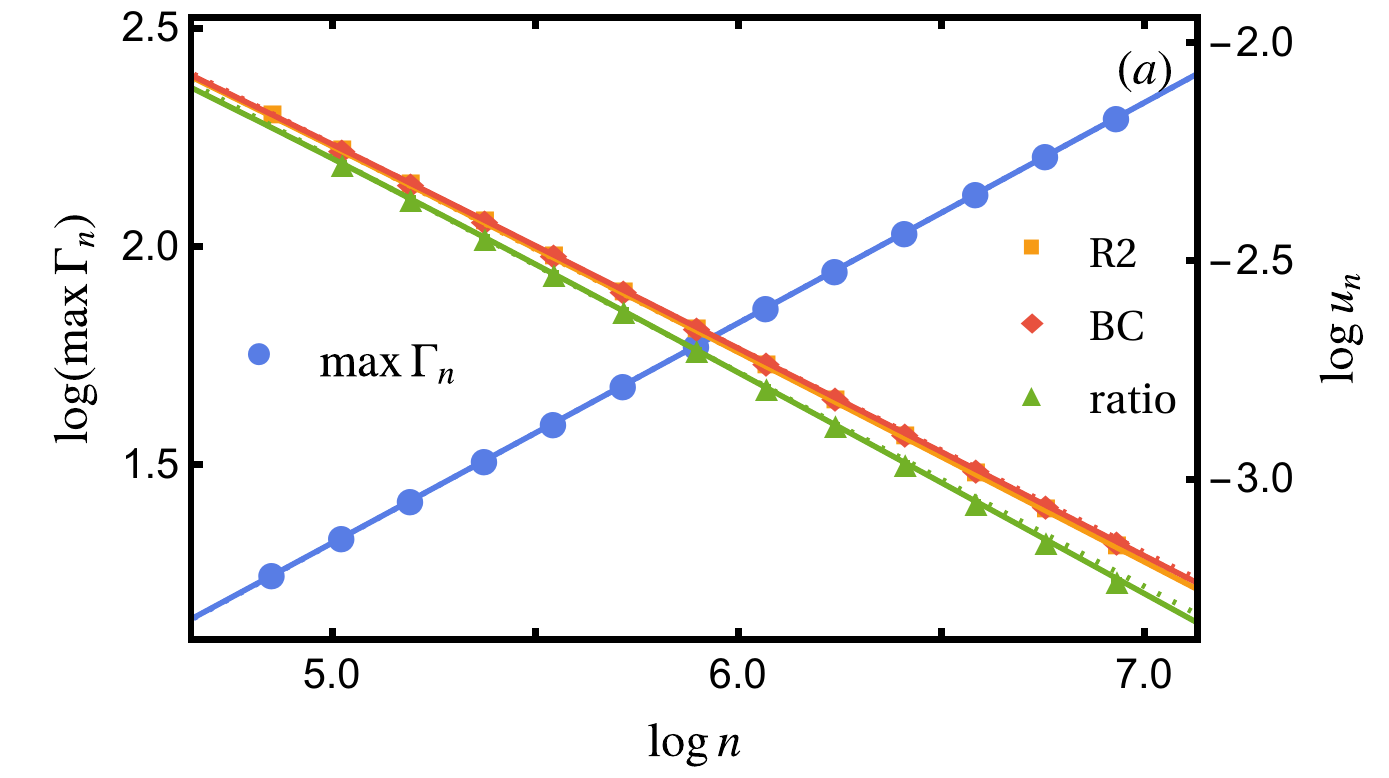}
	\includegraphics[width=\columnwidth]{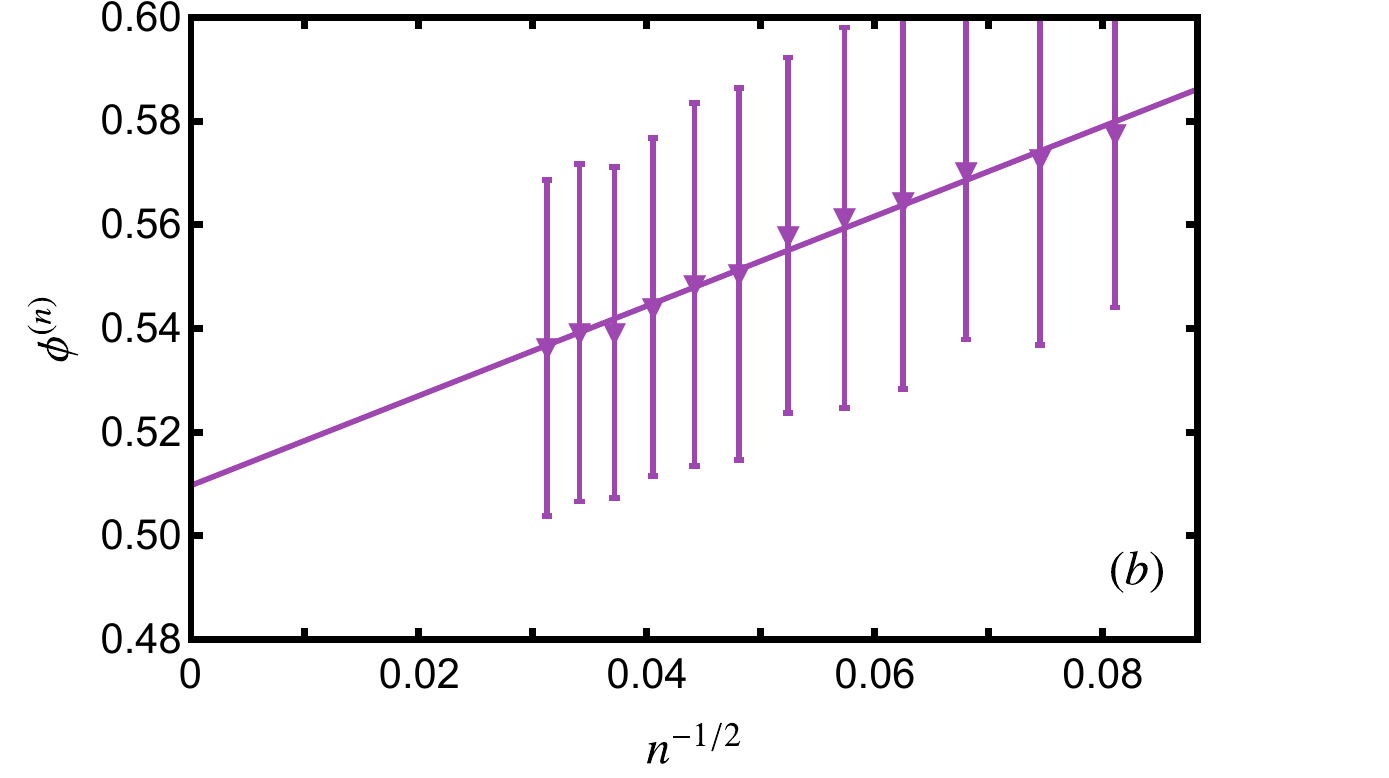}
	\includegraphics[width=\columnwidth]{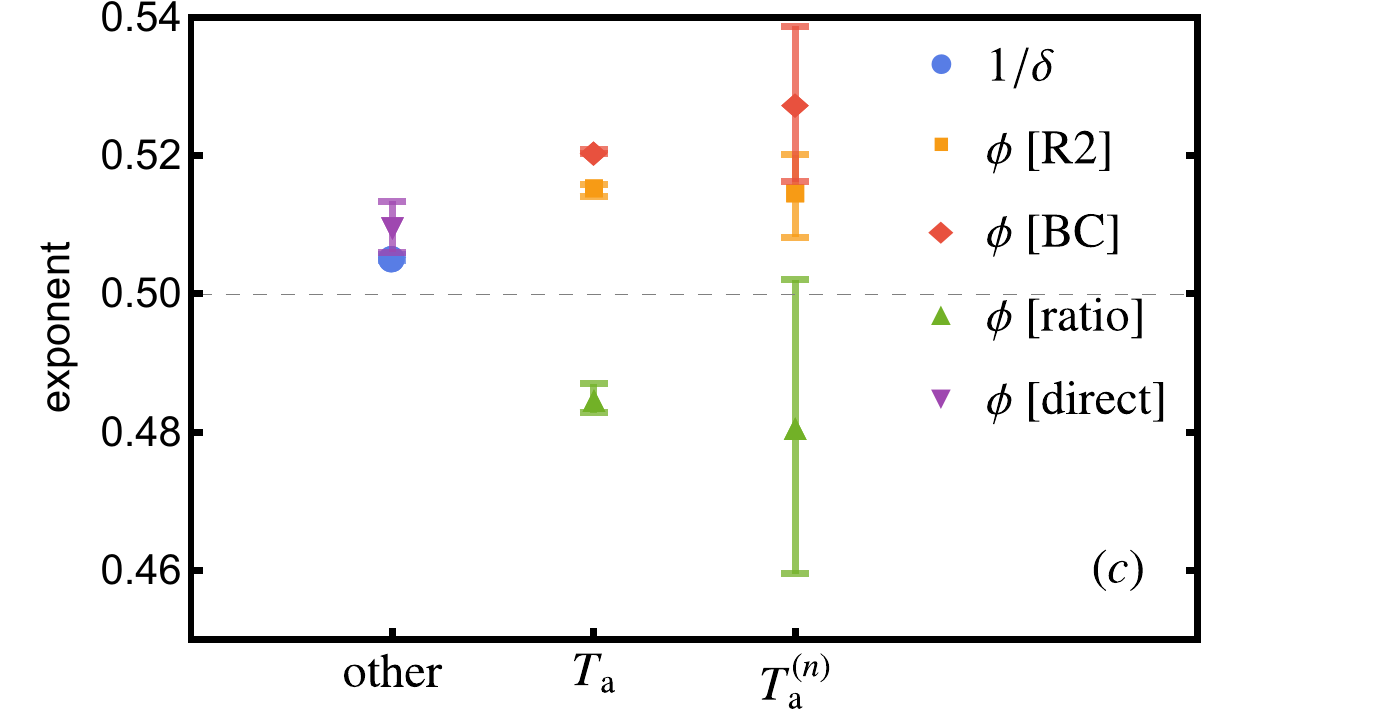}
	\caption{For SAWs on the square lattice, (a) log-log plot of $\Gamma_n$ vs.~$n$ and $u_n$ vs.~$n$. The latter are calculated using the extrapolated values of $\Tcinf$ from the BC, R2 and ratio methods. Solid curves are appropriate fits with correction-to-scaling term. Power-law only fits are also shown as dotted lines, where visible. (b) Plot of $\phi^{(n)}$ calculated directly from ratios of $u_n$, vs.~$1/\sqrt{n}$, and extrapolated to large $n$. (c) Estimates of the exponents using the various approaches discussed in the text. For specific values see Table \ref{tab:ExponentResultsBest}.}
	\label{fig:SquSAWsAnalysis}
	\vspace{-0.5cm}
\end{figure}

\begin{figure*}[t!]
	\centering
	\includegraphics[width=\textwidth,trim={0 1cm 0 1cm},clip]{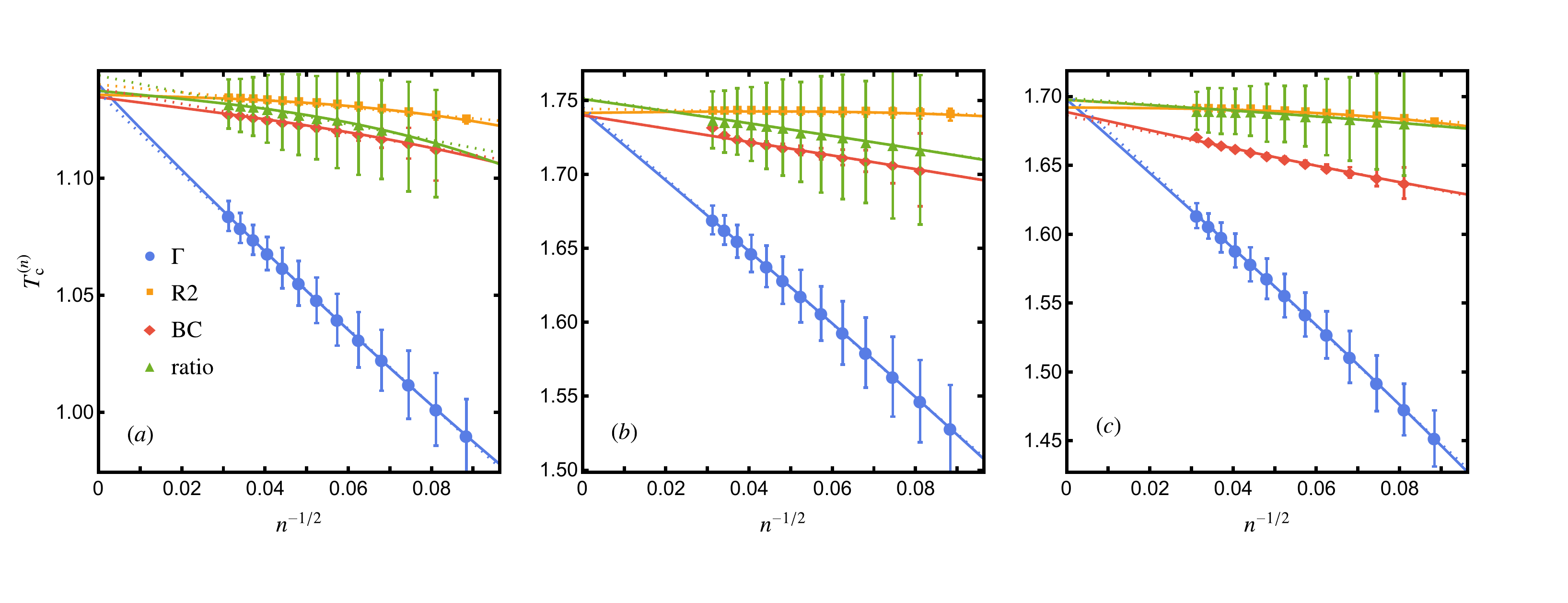}
	\caption{Finite-size critical temperatures for two-dimensional lattice models (a) SAWs on the hexagonal lattice and (b) SAWs and (c) SATs on the square lattice. For each of the four methods the solid lines are fits with correction to scaling and dotted lines are power-law only.}
	\label{fig:2DTemperatures}
	\vspace{-0.5cm}
\end{figure*}

As discussed in \sref{sec:Methods} and following \cite{Plascak2017}, the canonical method to estimate the exponents is as follows. The first step is to calculate $1/\delta$ from $\Gamma_n$. A log-log plot of $\max\Gamma_n$ for \smash{$n=128,\ldots,1024$} is shown in \fref{fig:SquSAWsAnalysis}(a) (blue, left) along with fits to \eref{eq:GammaScaling}. We get \smash{$1/\delta=0.5264(12)$} for a power-law only fit and \smash{$1/\delta=0.51528(86)$} by including a correction-to-scaling term. 
Although there is not a lot of difference between the fits on this scale, given the known value of \smash{$1/\delta=1/2$} in two dimensions, it is clear that the correction-to-scaling term is significant. 

The next step is to consider the critical temperatures. Figure \ref{fig:2DTemperatures}(b) shows the finite-size critical temperatures, $\Tcn$, for SAWs on the square lattice, using the four methods discussed in \sref{sec:Methods}. 
Using the (correction-to-scaling) value of $1/\delta$ just found, we also show fits according to \eref{eq:TempScaling} for each of these sets. Solid lines are fits with correction-to-scaling term, and dotted lines are power-law only. Extrapolating the power-law only fits to \smash{$n\to\infty$} obtains \smash{$\Tcinf=1.74548(70)$}, $1.74001(49)$, $1.74429(76)$, $1.7517(15)$ for the $\Gamma$, R2, BC and ratio methods, respectively. Using correction-to-scaling fits instead obtains \smash{$\Tcinf=1.74292(54)$}, $1.7399(19)$, $1.74151(69)$, $1.7510(57)$ for the $\Gamma$, BC, R2 and ratio methods, respectively.

The $\Tcinf$ from each method appear to have good agreement, yet a few points should be made. Firstly, although it may not be clear from just the reported values for the case of square lattice SAWs, the correction-to-scaling fits are generally better than using power-law only. The R2 method is the best for locating the $\Tcn$, having much less variation over this range of $n$, and having much smaller error bars for each $\Tcn$ than other methods. The small error bars are in due in part to the fact that the method relies on intersections of near-perpendicular curves, as opposed to the near-parallel curves of the ratio method. This more than counteracts the lack of correction-to-scaling terms in the R2 method compared to the rest of the analysis.

The BC method, via the Binder cumulant, presents the most difficulty. Notice that the $\Tcn$ deviate from the trend at large $n$. We note that the correction-to-scaling term cannot account for this kink and even if it could the extrapolation $n\to\infty$ would be significantly different from the other methods. Instead, we account for this by only using data up to $n\lesssim 600$ in the fits, where the scaling law fits well. This cutoff was determined to be the point where the error in the fitting parameters started to diverge as data for larger $n$ was added to the fit.

As to why this kink is present, we hypothesize that it is a limitation of finite simulations. While our data is equilibrated to a high degree, the fourth-order moment $\langle m^4\rangle$ that appears in \eref{eq:Binder} is more susceptible to error as $n$, and therefore maximum possible values of $m$, increase. It would take orders of magnitude more samples to ensure that fourth-order moments are equilibrated.
Of course, we cannot rule out that it is a quirk of the flatPERM algorithm and other simulation methods may not have this issue. However, we note that our simulation is up to the reasonably long length of $1024$ and the kink occurs at greater lengths than those considered in previous works that use the Binder cumulant \cite{Plascak2017}.

The ratio method of determining $\Tcinf$ also has some flaws. The individual $\Tcn$ are closer to the R2 method than the others, yet the individual error bars are much larger, and the extrapolated value $\Tcinf$ does not agree with the other three methods. However, the latter point is not a general observation for all lattice models.


Lastly, the $\Gamma$ method is interesting because at first glance the extrapolated value of $\Tcinf$ appears to agree with the other methods. This is in contrast to the obvious difference between this method and the others at finite $n$, as clearly visible in \fref{fig:2DTemperatures}(b). This gap is indicative of the fact that the locations of the peaks of $\Gamma_n$ are not claimed to properly approximate the critical temperatures. In fact, the scaling variable, \smash{$x=(\Tcinf-T)\,n^{1/\delta}$}, is significantly greater than unity for the $\Gamma$ method. Attempting to use this method anyway is fraught due to the relation to the specific heat, as mentioned earlier. It is also a distinctly different approach to the other methods which all have the common aspect that curves for different $n$ should intersect near the critical temperature, representing the existence of a universal value of the given thermodynamic quantity at the critical temperature. Furthermore, although not necessarily clear for SAWs on the square lattice, on closer inspection the values of $\Tcinf$ for the $\Gamma$ method are generally off compared to the other three methods. Given these concerns, we thus record the extrapolated value of the critical temperature for the $\Gamma$ method, but will not go on to use it to calculate $u_n$ and thus $\phi$. Even without this argument, the resulting values of $\phi$ are consistently off compared to the other three valid methods.

Turning to $\phi$, a log-log plot of $u_n(\Tcinf)$ for \smash{$n=128,\ldots,1024$} is shown in \fref{fig:SquSAWsAnalysis}(a) (right) along with fits to \eref{eq:UnScaling}. Since the $\Tcinf$ estimates are so close together for this lattice model the curves of $u_n$ overlap strongly on this scale. For the three valid finite-size scaling methods, BC, R2 and ratio, and using power-law only fits, this obtains \smash{$\phi=0.5325(17)$}, $0.5292(23)$ and $0.5094(35)$, respectively. Including the correction-to-scaling term gives \smash{$\phi=0.52062(24)$}, $0.51493(87)$ and $0.4849(21)$, respectively. Here, the correction to scaling is a clear improvement for the R2 method, marginal for the BC method and questionable for the ratio method.

There is also an alternative approach whereby we evaluate $u_n(\Tcn)$ at each different $\Tcn$, rather than the single extrapolated $\Tcinf$. This is similar to the calculation of $1/\delta$ where the maxima of $\Gamma_n$ occur at different temperatures for each $n$. Using the correction-to-scaling fits, this obtains {$\phi=0.527(11)$}, $0.5142(60)$ and $0.481(21)$. The choice of whether to use $\Tcinf$ or the set of $\Tcn$ is not {\em a priori} clear, but the resulting values of $\phi$ have much larger errors and spread between methods. They are shown in \fref{fig:SquSAWsAnalysis}(c) as a comparison, but it is clear that using the single $\Tcinf$ to obtain $\phi$ is a better approach.

The last estimate of $\phi$ comes from the ratio method which, as explained in \sref{sec:Methods}, estimates $\phi$ more directly by extrapolating the $\phi^{(n)}$ at the critical points to \smash{$n\to\infty$}, as shown in \fref{fig:SquSAWsAnalysis}(b). We assume the ansatz
\begin{equation}
	\phi^{(n)} = \phi + \frac{C}{\sqrt{n}} + \ldots,
	\label{eq:}
\end{equation}
where $C$ is a constant, obtaining \smash{$\phi=0.5097(37)$}. In total, we thus obtain four estimates of $\phi$ for SAWs on a square lattice - three from valid finite-size scaling methods and one from the direct method - and one estimate of $1/\delta$, All are listed in Table \ref{tab:ExponentResults} and we will discuss how to combine these values in the next section.

\subsection{Two dimensions}
\label{sec:2DResults}

\setlength{\tabcolsep}{2pt}
\begin{table}[t!]
	\caption{Valid results for all lattice models and methods. All values are from fits with correction-to-scaling terms. 
	}
	\begin{tabular}{l|l|llllr}
	\hline \hline
	 & Method & \multicolumn{1}{c}{$1/\delta$} & \multicolumn{1}{c}{$\Tcinf$} & \multicolumn{1}{c}{$\phi$} & \multicolumn{1}{c}{$\phi$ \,[$\Tcn$]} \\
	\hline
	
	hex  	& $\Gamma$ 	& $0.4851(11)$ & $1.14014(51)$ & $-$ 	& $-$ 			\\ 
	SAW		& BC 		& $-$ 	& $1.13465(40)$ & $0.51014(76)$ & $0.5137(55)$ 	\\
			& R2 		& $-$ 	& $1.13566(44)$ & $0.5058(13)$ 	& $0.5077(46)$ 	\\ 
			& ratio 	& $-$ 	& $1.1374(19)$ 	& $0.4960(16)$ 	& $0.499(12)$ 	\\ 
			& direct 	& $-$ 	& $-$ 			& $0.5002(20)$ 	& $-$ 			\\ [1ex]
			& fixed $\kappa$ 	& $0.5060(12)$ 	& $1.13459\ldots$ & $0.496(10)$ & $-$ \\ [1ex]

	squ		& $\Gamma$ 	& $0.50525(40)$ & $1.74292(51)$ & $-$ 	& $-$ 			\\ 
	SAW		& BC 		& $-$ 	& $1.7399(19)$ & $0.52062(24)$ & $0.527(11)$ 	\\
			& R2 		& $-$ 	& $1.74151(69)$ & $0.51493(87)$ & $0.5142(60)$ 	\\ 
			& ratio 	& $-$ 	& $1.7510(57)$ 	& $0.4849(21)$ 	& $0.481(21)$ 	\\
			& direct 	& $-$ 	& $-$ 			& $0.5097(37)$ 	& $-$ 			\\[1ex]
	
	squ		& $\Gamma$ 	& $0.50393(39)$ & $1.6978(14)$ & $-$ 	& $-$ 			\\ 
	SAT		& BC 		& $-$ 	& $1.6887(15)$ 	& $0.51172(67)$	& $0.516(12)$ 	\\
			& R2 		& $-$ 	& $1.69201(74)$ & $0.50127(17)$ & $0.5029(56)$ 	\\ 
			& ratio 	& $-$ 	& $1.6975(34)$ 	& $0.4839(10)$ 	& $0.482(10)$ 	\\
			& direct 	& $-$ 	& $-$ 			& $0.4973(23)$ 	& $-$ 			\\[1ex]

	sc		& $\Gamma$ 	& $0.47911(56)$ & $3.5504(73)$ & $-$ 	& $-$ 			\\ 
	SAW		& BC 		& $-$ 	& $3.5146(83)$ 	& $0.4887(19)$ 	& $0.500(15)$ 	\\
			& R2 		& $-$ 	& $3.5271(36)$ 	& $0.4799(24)$ 	& $0.4691(54)$ 	\\ 
			& ratio 	& $-$ 	& $3.519(20)$ 	& $0.4847(21)$ 	& $0.474(31)$ 	\\
			& direct 	& $-$ 	& $-$ 			& $0.4907(19)$ 	& $-$ 			\\[1ex]
	
	sc		& $\Gamma$ 	& $0.48368(40)$ & $3.7557(85)$ & $-$ 	& $-$ 			\\ 
	SAT		& BC 		& $-$ 	& $3.707(12)$ 	& $0.4927(12)$ 	& $0.493(14)$ 	\\
			& R2 		& $-$ 	& $3.7294(53)$ 	& $0.4745(25)$ 	& $0.4717(52)$ 	\\
			& ratio 	& $-$ 	& $3.726(11)$ 	& $0.4800(18)$ 	& $0.482(21)$ 	\\
			& direct 	& $-$ 	& $-$ 			& $0.4865(16)$ 	& $-$ 			\\
	\hline \hline
	\end{tabular}
	\label{tab:ExponentResults}
\end{table}


We now present the results for the other two-dimensional lattice models and discuss how to combine the results. For each lattice model, the intermediate quantities $\Gamma_n$, $u_n$ and $\phi^{(n)}$ and much of the details of the calculations are qualitatively identical to that of square SAWs discussed in the preceding section. In fact, the results of square SAWs tend to have larger errors and some of the issues are less of a problem in the other lattice models. As such we skip to presenting temperature and exponent results for the other cases. Furthermore, we also saw in the last section that fitting to the scaling forms is generally improved by the addition of a correction-to-scaling term and this is more true for the other lattice models. Henceforth we report only the correction-to-scaling results, where applicable.  

\begin{figure}[t!]
	\centering
	\includegraphics[width=\columnwidth]{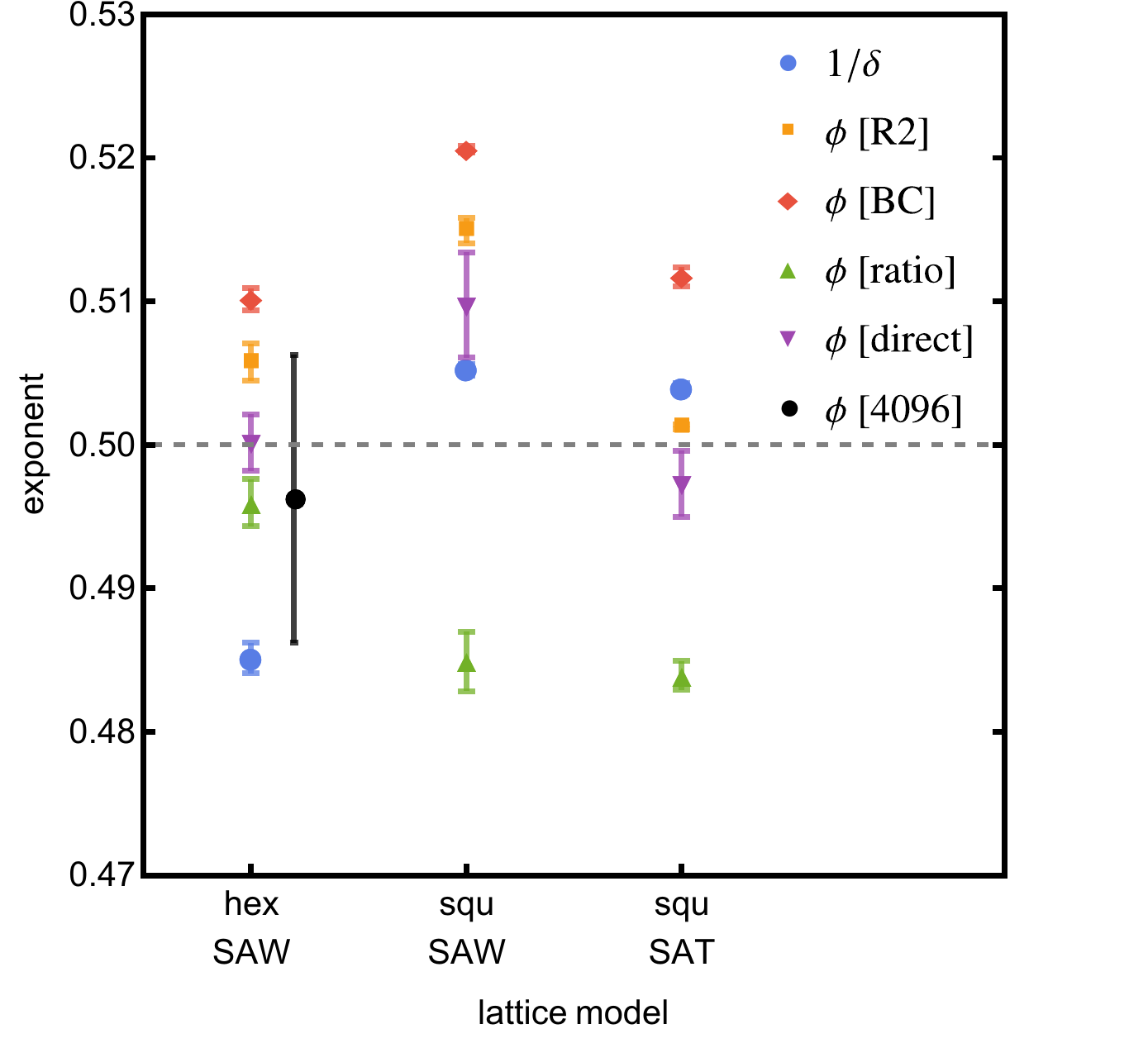}
	\caption{Exponents for two-dimensional simulations. Black is the special case of SAWs on hexagonal lattice simulated at fixed exact critical temperature up to \smash{$n=4096$}. The dashed gray line marks the expected value of the 2D crossover exponent $\phi=1/2$.}
	\label{fig:2DExponents}
	\vspace{-0.5cm}
\end{figure}

Figure \ref{fig:2DTemperatures} shows the critical temperatures for the two-dimensional lattice models. The results of extrapolating the fits to $\Tcinf$ are reported in Table \ref{tab:ExponentResults}, along with all estimates of exponents $1/\delta$ and $\phi$. Additionally, we visualize the exponents in Figures \ref{fig:2DExponents}. 
For these plots, the horizontal axis has no meaning except to cluster the results for each lattice model.

In addition to all the methods discussed so far, for the case of SAWs on the hexagonal lattice, we have the further benefit of knowing the exact critical temperature \smash{$\kcinf=1+\sqrt{2}$} \cite{Batchelor1995}. Incorporating this weight directly into the simulation greatly reduces equilibration time and allowed us to simulate SAWs on the hexagonal lattice up to length \smash{$n=4096$} in the same time as the full simulations up to length $1024$. In this case we do not need to locate the finite-size critical temperatures $\Tcn$; the exponents are determined directly from $\Gamma_n$ and $u_n$, obtaining \smash{$1/\delta=0.5060(12)$} and \smash{$\phi=0.496(10)$}. Note that the former comes from the scaling of $\Gamma_n(\kappa_c)$ rather than $\max\Gamma_n$; an inverse of the $\Gamma$ method for the other lattice models, potentially with similar limitations to estimating $1/\delta$. However, the value of $\phi$ is shown in \fref{fig:2DExponents} (black) for comparison to other lattice models and methods. Despite the ability to simulate much larger chains, the statistics of this simulation are not the same as the others and so these values should be considered a benchmark only. Nevertheless, it is a good test of the accuracy of the flatPERM algorithm and the significance of corrections to scaling in our methods. It also validates using $\Tcinf$ over the set of $\Tcn$. 

Regarding the critical temperatures, we immediately see that several features mentioned in the analysis of square SAWs are common to all lattice models. As mentioned earlier, the $\Gamma$ method is the worst at estimating the critical temperature at finite lengths, and is known to be an unreliable method. So, while the temperatures from the $\Gamma$ method have been shown, this method is not used in any further results. The R2 method appears to be the best at locating the critical temperature, given that the errors in $\Tcn$ are the smallest for this method, and the trend as \smash{$n\to\infty$} displays the smallest correction to finite-size scaling. The values of $\Tcn$ from the ratio method are very close to those of the R2 method for most lattice models, yet the errors are much larger due to the way curves of $\phi^\text{(n)}$ intersect. Nevertheless, resulting $\Tcinf$ and $\phi$ estimates from the ratio method are good.

One exception is for the BC method on the hexagonal lattice, where the deviation from trend at larger $n$ does not occur like the square lattice models. however, for consistency, we make the same restriction to \smash{$n\lesssim 600$}. A more general issue with the BC method is that it is parameter dependent, namely due to the minimum value of $n$ used as the common interceptor with curves of $U_4$ at larger $n$. We use \smash{$n=128$} as the minimum, intending that the range of $n$ is consistent with other methods, and thus the finite-size temperatures for the BC method are comparable to the R2 and ratio methods. If a larger range of $n$ is considered by using a smaller value for the minimum, then the temperatures are much closer to the $\Gamma$ method, which we have already noted as unreliable. We find our range of $n$ to be a good tradeoff between minimizing the effect of corrections to scaling from smaller $n$ and having enough data to achieve a good fit to the scaling form. Note that altering the minimum value of $n$ does not alter the value of $n$ where the kink starts.

Despite these cautions, it is not plausible to conclude that, say, the R2 method is better than the others and should always be used for these kind of calculations. Even if it appears to be the best way to determine $\Tcinf$ it is not overwhelmingly better than the other methods. The issues with the BC and ratio methods are technical, and should be retained as valid.
Thus, despite omitting some methods as invalid or too imprecise, we still have a spread in the valid estimates for the exponents, as seen in \fref{fig:2DExponents}. Rather than relying on the statistical errors reported so far, we instead view the variance in results as evidence of a larger systematic error.

Regarding the different exponents, it is further clear from \fref{fig:2DExponents} that $1/\delta$ falls within the spread of the $\phi$ estimates. One can compare $1/\delta$ to $\phi$ for a specific method to find a pattern, or omit certain values that appear to be outliers or unreliable, but generally, across all lattice models, this does not hold. We are forced to consider that $1/\delta$ is not distinct from $\phi$.
Moreover, we could even say that the calculation of $1/\delta$ from the scaling of $\max\Gamma_n$ is yet another method for estimating the crossover exponent of the adsorption transition, equally valid as using the three finite-size scaling methods or calculating $\phi$ directly from ratios of $u_n$.

Arguably, the main reason that the statistical error is so small is that it arises from the very small errors in the calculated thermodynamic functions which are in turn due to being averaged over the ten independent simulations for each lattice. 
The better statistics of the shorter length simulations cover the fact that we are not able to find the critical temperature as accurately due to correction-to-scaling effects and the differences in methodologies. Compare this to the simulation of \smash{$n=4096$} SAWs on the hexagonal lattice, where the error in $\phi$ is also statistical, but we have complete confidence in knowing the exact temperature. It is therefore striking that the black error bar in \fref{fig:2DExponents} is so comparable in magnitude to the spread of exponent estimates for the \smash{$n=1024$} simulations. Given the issues with reported statistical errors, when making these averages we omit statistical errors beyond the third decimal place as being too far removed from the systematic spread in values.

The final task is therefore to combine the results for all lattice models into results for two dimension, which we will apply to three dimensions in the next section. While it is not possible to pick out one method over another, we note that they are not all equivalent. The R2, BC and ratio methods are similar in that they first estimate the critical temperature which is then used to find $\phi$ from the finite-size scaling of $u_n$. In order to compare to $1/\delta$ and the direct estimates of $\phi$ we average the values of $\phi$ for the three finite-size scaling methods, obtaining $\phi^\text{(FSS)}$, listed in Table \ref{tab:ExponentResultsBest} for each lattice model. Also listed are the critical temperatures of each lattice model, averaged from the three finite-size scaling methods. 

\begin{figure*}[t!]
	\centering
	\includegraphics[width=0.75\textwidth,trim={0 1.5cm 0 1cm},clip]{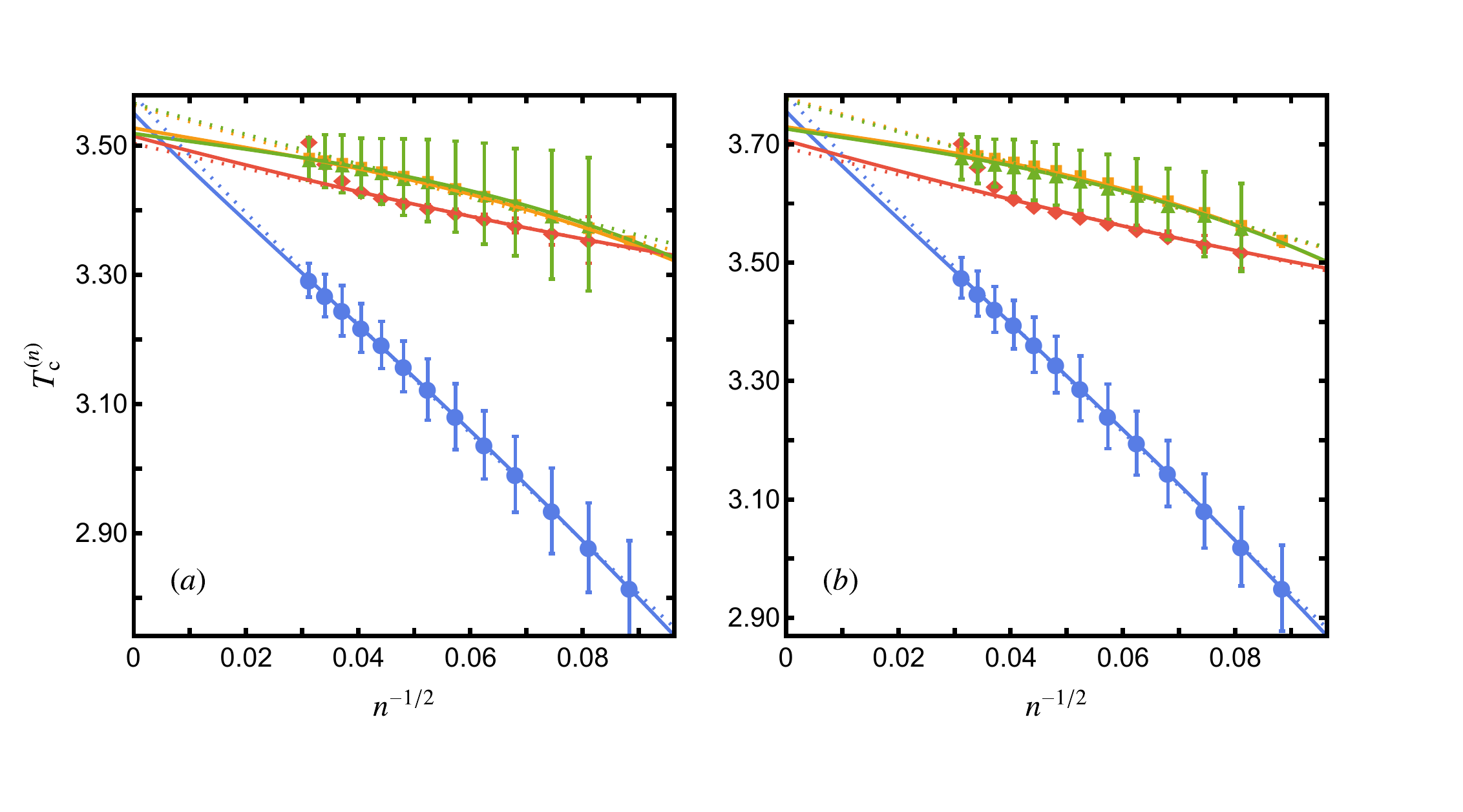}
	\caption{Finite-size critical temperatures for three-dimensional lattice models (a) SAWs and (b) SATs on the simple cubic lattice. For each of the four methods the solid lines are fits with correction to scaling and dotted lines are power-law only.}
	\label{fig:3DTemperatures}
	\vspace{-0.5cm}
\end{figure*}

\begin{figure}[b!]
	\centering
	\includegraphics[width=\columnwidth]{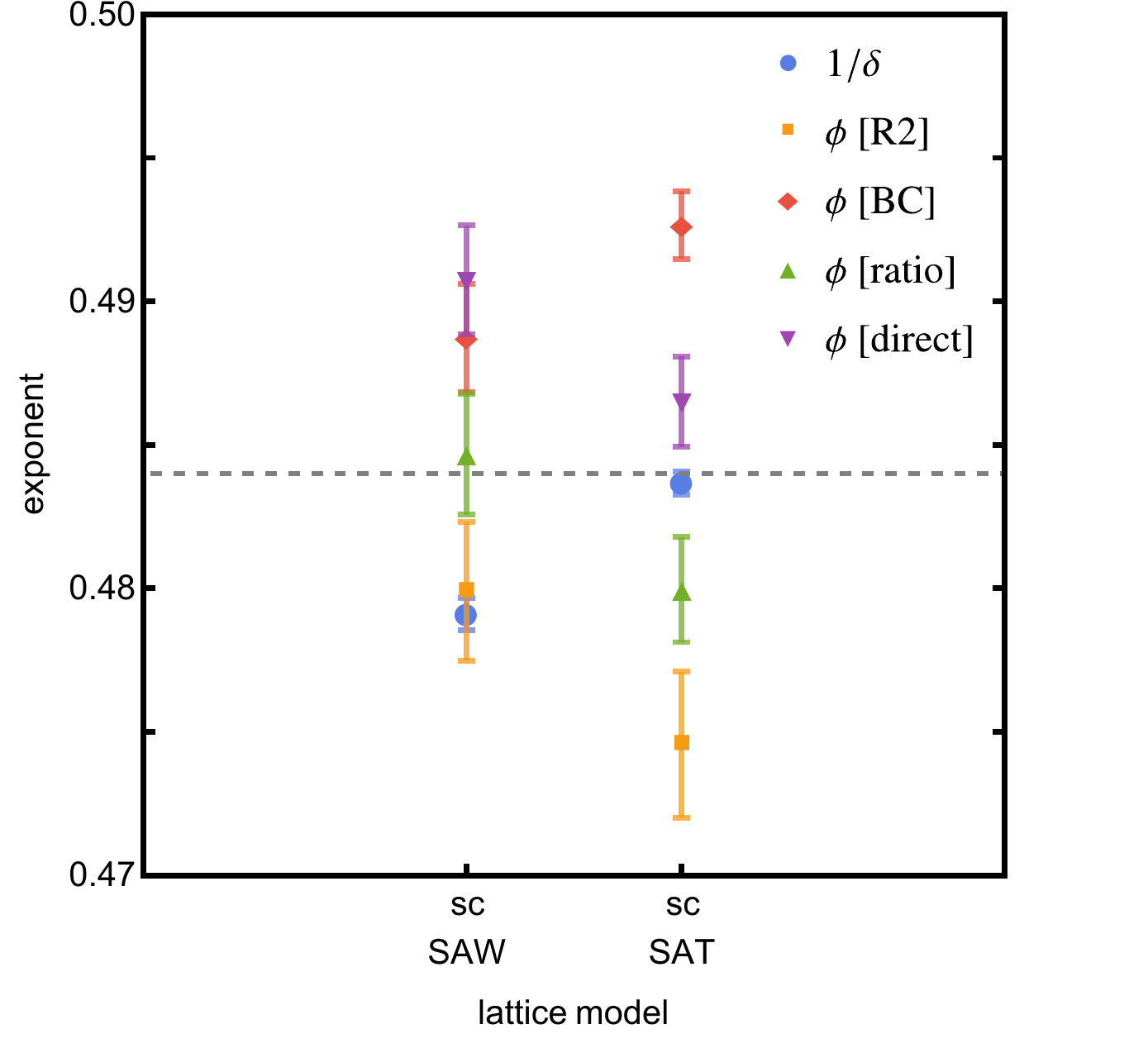}
	\caption{Exponents for three-dimensional (simple cubic) simulations. The dashed gray line marks the average estimate of the 3D crossover exponent $\phi=0.484(4)$.}
	\label{fig:3DExponents}
	\vspace{-0.5cm}
\end{figure}

The $\phi^\text{(FSS)}$ value is now comparable to $\phi^\text{(direct)}$, which is from $u_n$ but without finite-size-scaling, and to $1/\delta$ which comes from a different, but related thermodynamic quantity. We average these three values equally to obtain the exponent for each lattice model, also listed in Table \ref{tab:ExponentResultsBest}. Recall that the values for $1/\delta$ and the direct estimate for $\phi$ are already listed in Table \ref{tab:ExponentResults}. Finally, the exponents are averaged over all lattice models in each dimension. Thus, for two-dimensions we obtain \smash{$\phi=0.501(2)$}, in agreement with the known value of $1/2$. The uncertainty in this value is due to the spread from the different methods and models rather than the statistical error in those values. There is still some spread in the value of this final exponent for the two-dimensional lattice models, 

As a final remark, we note that an alternative approach is to use the average $\Tcinf$ to calculate a single $\phi^\text{(FSS)}$, but we found no meaningful difference. It is well known that the value of $\phi$ is sensitive to accurately knowing the critical temperature. This alternative would require an estimate of the error in $\phi$ by propagating the error in the average temperature, itself a product of the spread in individual values $\Tcinf$. By \eref{eq:UnScaling}, this is not a straightforward procedure. We found that any reasonable attempt to do this produces an error in $\phi$ that is the same magnitude as the spread in $\phi$ values from individual methods as already reported. Thus the presence of a systematic error is clear either way.

\subsection{Three dimensions}
\label{sec:3DResults}

Having verified our methodology on the two-dimensional lattice models, we now turn to the three-dimensional simple cubic lattice models. The critical temperatures for SAWs and SATs on the simple cubic lattice are shown in \fref{fig:3DTemperatures} and the exponents are visualized in \fref{fig:3DExponents} and listed in Table \ref{tab:ExponentResults}.

The main point where the analysis of the simple cubic lattice models differs from the two-dimensional cases is with the kink in critical temperatures from the BC method. In \fref{fig:3DTemperatures} we see that at higher $n$ the $\Tcn$ diverge faster than for the square lattice. However, the point at which this kink begins is the same, so we have the same range of $n\lesssim 600$ for this method. All other methods proceed in the same manner as in the two-dimensional analysis.

The final results for the critical temperatures and exponents for the simple cubic lattice models are determined in the same way as the two-dimensional case and are summarized in Table \ref{tab:ExponentResultsBest}.
For SAWs on the simple cubic lattice we find, after averaging over the different finite-size scaling methods, that \smash{$\phi^\text{(FSS)}=0.484(4)$}, compared with \smash{$\phi^\text{(direct)}=0.491(2)$} and \smash{$1/\delta=0.4791(6)$}. Similarly, for SATs we find \smash{$\phi^\text{(FSS)}=0.482(9)$}, compared with \smash{$\phi^\text{(direct)}=0.487(2)$} and \smash{$1/\delta=0.4837(4)$}. As with the two-dimensional case, and knowing the source of the error bars, these values are not distinct enough to definitively separate them. Hence, assuming equality of $\phi$ and $1/\delta$, we estimate \smash{$\phi=0.485(6)$} for SAWs and \smash{$\phi=0.484(2)$} for SATs.

Averaging over the values of both three-dimensional models gives our best estimate \smash{$\phi=0.484(4)$}. Even given the magnitude of the potential systematic error, we conclude that for three-dimensions $\phi$ does deviate from the mean-field value of $1/2$. However, we find that there is not a clear difference between SAWs and SATs, nor do we find evidence for $1/\delta$ being different from $\phi$.


\begin{table}[t!]
	\caption{Best results for the adsorption temperature and the finite-size scaling estimates of $\phi$ for each lattice model. Bold values are the combined result for the crossover exponent for each lattice model and dimension.}	
	\begin{tabular}{l|ll|l}
	\hline \hline
	 & \multicolumn{1}{c}{$T_c$} 
	& \multicolumn{1}{c}{FSS $\phi$} & \multicolumn{1}{c}{$\phi$ [$=1/\delta$]}\\\hline	
				
	hex~4096 & 1.13459\ldots& 0.496(10) &\\  \hline
	hex~SAW		& 1.136(1)	& 0.504(7) 	& {\bf 0.496(10)} \\ 
	squ~SAW	   	& 1.744(6) 	& 0.507(19) & {\bf 0.507(2)} \\ 	
	squ~SAT		& 1.693(4) 	& 0.499(14) & {\bf 0.500(3)} \\ \hline
	2D 			&	  		&  			& {\bf 0.501(2)} \\ 
	\hline

	sc~SAW		& 3.520(6) 	& 0.484(4)	& {\bf 0.485(6)} \\
	sc~SAT		& 3.720(12)	& 0.482(9) 	& {\bf 0.484(2)} \\ \hline
	3D 			& 			& 			& {\bf 0.484(4)} \\
	\hline\hline
	\end{tabular}
	
				%
	%
	\label{tab:ExponentResultsBest}
\end{table}



\section{Conclusion}
\label{sec:Conc}


We have performed a comprehensive study of self-avoiding walks and trails on two- and three-dimensional lattices with an adsorbing boundary. Numerical simulations up to polymer length of 1024 provide a wealth of data for studying the adsorption transition. A variety of analyzes were used to estimate the critical temperature and scaling exponents of this transition. Using both the square and hexagonal lattices, and in the latter case also using exact results, we confirm the mean-field value of the crossover exponent \smash{$\phi=1/2$} (also obtained from exact solution methods and conformal field theory), with our own estimate of \smash{$\phi=0.501(2)$}. What is not apparent in this final result is that applying individually valid methods to each of the lattice models produces a large spread in estimates. This suggests a \emph{significant systematic error} in any individual estimate greater than the statistical error intrinsic to the numerical analysis.

Applying the same methodology, averaging over several estimates, to the three-dimensional lattice models of SAWs and SATs on the simple cubic lattice, we provide a final estimate \smash{$\phi=0.484(4)$}. This is in agreement with other recent works that suggest a deviation from the mean-field value in three dimensions, and thus that the crossover exponent is not super-universal. However, as with two dimensions, there is systematic error across the different methodologies which does not allow for a distinction between the crossover exponent $\phi$ and the shift exponent $1/\delta$. In fact, we suggest that direct estimates of the shift exponent are yet another way of estimating the crossover exponent, and not of estimating a distinct quantity.

As well as variety in the analysis of thermodynamic quantities, we have considered walks and trails equally for the square and simple cubic lattices. As the SAW model can be considered as the strongly repulsive limit of the interacting SAT model, the agreement of our results for the two models also indicates that the universality of the critical exponent is not broken by (repulsive) monomer-monomer interactions. Of course, this constitutes only two data points on the scale of variable monomer-monomer interaction strength but assuming universality raises the possibility of more accurate exponent estimates:  Considering both the interacting walk and interacting trail models one may be able to achieve greater accuracy in locating the critical temperature by varying the monomer-monomer interactions to minimize corrections-to-scaling in quantities such as the end-to-end distance scaling, which has had some previous success \cite{Prellberg2001}. Additionally, it raises the question of studying the general interacting SAT model with strongly attractive interactions, known to be in a different universality class to the interacting SAW model at its collapse point. 


\begin{acknowledgments}
Financial support from the Australian Research
Council via its Discovery Projects scheme (DP160103562)
is gratefully acknowledged by the authors. Numerical simulations were performed using the HPC cluster at University of Melbourne (2017) Spartan HPC-Cloud Hybrid: Delivering Performance and Flexibility. https://doi.org/10.4225/49/58ead90dceaaa
\end{acknowledgments}

\bibliography{polymers}{}

\end{document}